\documentclass[a4paper, 11pt]{article}

\usepackage{float}
\usepackage{lineno}  
\usepackage{amsmath, amssymb}  
\usepackage{graphicx}  
\usepackage{geometry}  
\usepackage{authblk} 
\usepackage[hidelinks]{hyperref}  
\usepackage{xspace}
\usepackage{todonotes}
\usepackage{xparse}
\usepackage{rotating}
\usepackage{afterpage}

\newcommand{\orcid}[1]{\href{https://orcid.org/#1}{ORCID: #1}}

\newcommand{\figref}[2]{\textbf{Fig.~\ref{#1}#2}\xspace}

\newcommand{\tableref}[1]{\textbf{Table~\ref{#1}}\xspace}

\usepackage[backend=biber,style=nature, doi=true, url=false, isbn=false, eprint=false]{biblatex}
\AtEveryBibitem{%
  \clearlist{language}
  \clearfield{langid}
  \iffieldundef{doi}
    {\ifentrytype{online}{}{
      \printfield{url}%
      \iffieldundef{urldate}{}{%
        \printtext[parens]{\printurldate}%
      }%
    }%
    }
    {
      \clearfield{url}
      \clearfield{urldate}
      \clearfield{date}
      \clearfield{note}
      \clearfield{month}
    }
}

\newcounter{extendeddatafigure}

\newcommand{\extendeddatafigures}{
    \setcounter{extendeddatafigure}{0} 
    \renewcommand{\figurename}{} 
}

\newenvironment{extendeddatafiguresenv}{
}{
}

\newenvironment{extendeddatafigure}[1][]{
    \refstepcounter{extendeddatafigure} 
    \begin{figure}[#1]
     
}{
    \end{figure}
}

\NewDocumentCommand{\extref}{m o}{
    \hyperref[#1]{\textbf{\ref*{#1}}}
    \IfValueT{#2}{\textbf{#2}}%
}

\newcommand{\beginsupplement}{%
    \setcounter{table}{0}%
    \renewcommand{\thetable}{S\arabic{table}}
    \renewcommand{\tablename}{Supplementary Table}
}

\NewDocumentCommand{\suppref}{m o}{
    \textbf{\tablename~\ref{#1}}
    \IfValueT{#2}{\textbf{#2}}
}

\begin{document}

\title{The Reproducible Research Platform establishes a unified open science environment bridging data and software lifecycles across disciplines, from proposal to publication}

\author[1,2,*]{Andreas P. Cuny \orcid{0000-0001-9640-4154}} 
\author[3]{Henry L\"utcke \orcid{0000-0001-7225-4374}} 
\author[3]{Andrei-Valentin Plamad\u{a} \orcid{0000-0003-3284-8869}} 
\author[2,3]{Antti Luomi}
\author[2,3]{John Hennig}
\author[2,3]{Matthew Baker}
\author[1,2]{Fabian Rudolf}
\author[2,3,*]{Bernd Rinn}

\affil[1]{Computational Systems Biology, ETH Zurich D-BSSE, Klingelbergstrasse 48, 4056 Basel, CH}
\affil[2]{Swiss Institute of Bioinformatics, Klingelbergstrasse 48, 4056 Basel, CH}
\affil[3]{ETH Zurich Scientific IT Services, 8092 Zurich, Switzerland}
\affil[*]{Correspondence to: cunya@ethz.ch or brinn@ethz.ch}

\maketitle

\begin{abstract}

Many research groups aspire to make data and code FAIR and reproducible, yet struggle because the data and code life cycles are disconnected, executable environments are often missing from published work, and technical skill requirements hinder adoption. Existing approaches rarely enable researchers to keep using their preferred tools or support seamless execution across domains.
To close this gap, we developed the open-source Reproducible Research Platform (RRP), which unifies research data management with version-controlled, containerized computational environments in modular, shareable projects. RRP enables anyone to execute, reuse, and publish fully documented, FAIR research workflows without manual retrieval or platform-specific setup.
We demonstrate RRP’s impact by reproducing results from diverse published studies, including work over a decade old, showing sustained reproducibility and usability. With a minimal graphical interface focused on core tasks, modular tool installation, and compatibility with institutional servers or local computers, RRP makes reproducible science broadly accessible across scientific domains.

\end{abstract}

\section{Introduction}

Traditional approaches to managing, storing, analyzing, and publishing scientific data are increasingly inadequate for the scale and complexity of contemporary datasets and their processing \autocite{mani_genomics_2025}. The absence of standardized, interoperable tools undermines reproducibility, transparency, and data reuse, the core tenets of open science, and affects researchers at all levels, from students to senior investigators, who must collaborate efficiently despite widely varying technical expertise. Addressing these challenges is essential not only for individual projects but also for enabling collective verification and extension of published findings \autocite{huch_reusable_2025}.\\
Diverse domains generate large and heterogeneous datasets, including imaging and microscopy \autocite{ruan_image_2024, cuny_cell_2022, amat_fast_2014}, structural biology and crystallography\autocite{ke_convolutional_2018}, electrophysiology and biophysics\autocite{cuny_high-resolution_2022}, systems biology and omics\autocite{uyar_flexynesis_2025, poulos_strategies_2020, chen_giotto_2025, poldrack_long-term_2015}. 
Increasingly complex analytical workflows, encompassing preprocessing, quantification, statistical analysis, and visualization, demand precise documentation \autocite{poulos_strategies_2020, pepke_computation_2009}. Yet overreliance on conventional Materials and Methods sections, coupled with incomplete or inaccessible records, continues to impede reproducibility \autocite{seibold_computational_2021, marques_imaging_2020, committee_on_reproducibility_and_replicability_in_science_reproducibility_2019, kitzes_practice_2017}.\\
Data and code availability upon request does not reliably ensure access, reuse, or reproducibility. Empirical analyses of randomly selected Science articles showed only 44\% make data available, and just 26\% of findings could be reproduced\autocite{stodden_empirical_2018}. Missing information about computational environments further compromises reproducibility: a single outdated or unavailable library can render a published workflow unusable \autocite{cuny_live_2022, hatton_computational_2019, benureau_re-run_2018, cohen-boulakia_scientific_2017}. These barriers undermine both the repeatability of experiments and collaboration among researchers  with differing skill levels using shared code, data, and workflows.\\
The research community has sought to address these issues by promoting transparency and the FAIR data principles (Findable, Accessible, Interoperable, Reusable)\autocite{wilkinson_fair_2016, barker_introducing_2022}. Funding agencies and publishers increasingly mandate sharing of raw data and analysis code via domain-specific repositories or workflow systems\autocite{pampel_re3data_2023, semmelrock_reproducibility_2023, mcdermott_reproducibility_2021, boehm_quarep-limi_2021, pommier_applying_2019, stodden_enhancing_2016}. In computational sciences, reproducibility efforts rely on version control, workflow automation, and interactive notebooks that combine narrative and code for transparent reporting\autocite{knuth_literate_1984, benureau_re-run_2018, rule_exploration_2018, hatton_computational_2019}. Platforms like Binder\autocite{jupyter_binder_2018, beg_using_2021}, CodeOcean\autocite{clyburne-sherin_computational_2019}, and Google Colab\autocite{bisong_building_2019, carneiro_performance_2018, luppichini_cleverriver_2023} provide cloud execution of code or live computational environments\autocite{szabo_sage_2017, perkel_why_2018, kurtzer_singularity_2017, noauthor_renku_nodate, da_veiga_leprevost_biocontainers_2017, the_galaxy_community_galaxy_2022} but depend on lightweight repositories, which lack robust data management and compliance capabilities for institutional-scale research.\\
A Research Data Management System (RDMS) is therefore essential because tools like Binder\autocite{jupyter_binder_2018} depend on Git, unsuitable for large datasets. While Git-based tools offer provenance tracking for software, they lack integrated data, sample, and inventory management, workflow automation, and compliance functions of a full RDMS\autocite{hart_ten_2016, barillari_openbis_2016}. Even with FAIR data sharing and computational notebooks, researchers must still spend time retrieving data from multiple sources, resolving dependency conflicts, and aligning code, metadata, and environments across incompatible systems, tasks that often deter reproducibility \autocite{kitzes_practice_2017, mitra-behura_singularity_2022, benureau_re-run_2018}.\\
A research project typically begins with a research question, ideally organizing data and code for reuse and reproducibility by default. In practice, data and code life cycles often run independently (\figref{fig.rrp_overview}{a}) \autocite{courbebaisse_research_2023, bossaller_research_2023}. The data life cycle begins with research planning and data management plans (DMPs) defining strategies for collection, documentation, storage, analysis, publication, preservation, and reuse \autocite{michener_ten_2015, hart_ten_2016, committee_on_reproducibility_and_replicability_in_science_reproducibility_2019, noauthor_research_nodate}. Concurrently, the code life cycle involves developing scientific software and analysis pipelines that demand careful recording of dependencies, environments, execution contexts, and rigorous testing, tasks often neglected or technically challenging \autocite{courbebaisse_research_2023, crusoe_methods_2022, mitra-behura_singularity_2022, benureau_re-run_2018}. Despite their shared origin, managing data and code separately fragments workflows, loses provenance, and creates barriers to reporting, collaboration, and reproducibility.\\
Existing RDMS, including electronic lab notebooks (ELNs), laboratory information management systems (LIMS, https://www.limswiki.org) \autocite{eichinski_datatrack_2016, amorim_comparison_2017, barillari_openbis_2016, bauch_openbis_2011}, data archives\autocite{peroni_packaging_2022}, and research repositories\autocite{pampel_re3data_2023, sicilia_community_2017, thelwall_figshare_2016}, provide persistent identifiers and open access, but typically omit executable computational environments \autocite{szabo_sage_2017, perkel_why_2018, kurtzer_singularity_2017}. Likewise, platforms like GitHub and Google Drive do not readily accommodate large datasets, integrate with institutional RDM workflows, or ensure policy compliance. No current solution unifies robust data and metadata management with executable environments accessible to researchers of diverse technical backgrounds, underscoring the need for open-source infrastructure supporting reproducible, replicable research \autocite{committee_on_reproducibility_and_replicability_in_science_reproducibility_2019}.\\
To address these challenges, we propose the Reproducible Research Platform (RRP), integrating open data practices, executable computational environments, and institutional-grade RDM into a single platform enabling FAIR-by-design reproducibility across the entire research life cycle, bridging usability gaps between technically advanced and novice researchers. By unifying data, metadata, code, and computational context, RRP transforms static FAIR principles into living, executable research objects that anyone can inspect, reuse, and rerun. Through three concrete examples, we demonstrate how published results \autocite{cuny_pypocquant_2021, clarkson_archaeology_2015, beam_data-driven_2021} can be reproduced using RRP, reflecting diverse user habits and needs and showing how RRP streamlines publication workflows, including this manuscript. Widespread adoption of RRP could make reproducible, transparent, and collaborative research the default rather than an aspiration in computational science.

\section{Results}

\subsection{RRP bridges data and code life cycles in RDM}

\begin{figure*}[ht!]
	\centering
      \includegraphics[width=16cm ]{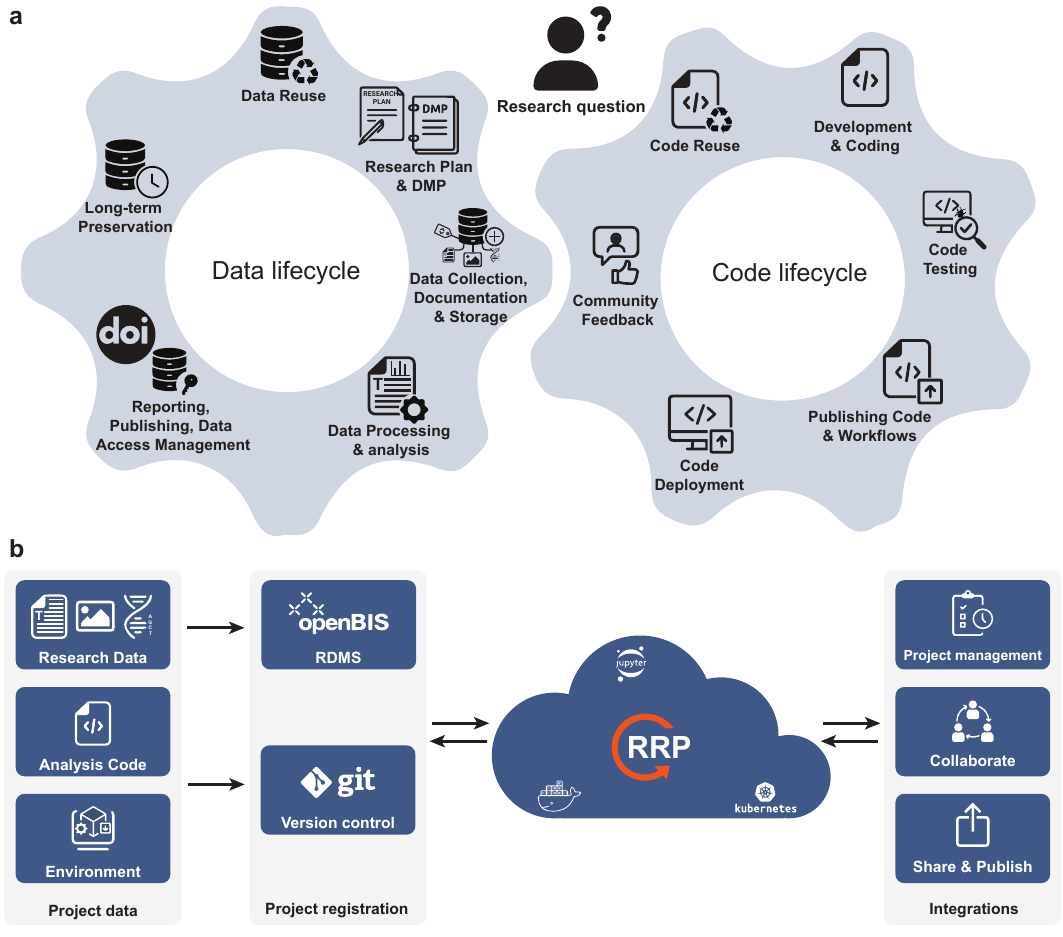}
	\caption{Data and code lifecycle of research projects and overview of RRP, an integrated, reproducible, and executable data storage and analysis platform. \textbf{a} Typical data and code lifecycles, often considered as separate processes, although beginning with a common research question and potential data and code reuse with key steps until publication and community engagement. \textbf{b} Research management within RRP, integrating both data and code within one platform. Research data are managed using the research data management system (RDMS), openBIS ELN-LIMS. Analysis scripts, code, and computational environment definitions are managed with Git (using platforms like GitLab or GitHub). With RRP, reproducible computational environments are created as Docker containers from the code repository using the open-source tool repo2docker. Users can access computational environments through popular user interfaces like JupyterLab. RRP directly mounts the relevant data stored in the RDMS. RRP directly enables the sharing of self-contained, executable projects facilitating collaborations with colleagues within the research group or with researchers worldwide. Overall, RRP accelerates research project management from planning until publication and beyond.}
	\label{fig.rrp_overview}
\end{figure*}

We developed the Reproducible Research Platform (RRP) to bridge a critical gap in research project management by unifying the traditionally separate data and code life cycles into a cohesive, project-centric environment that fully integrates data, code, metadata, and executable environments (\figref{fig.rrp_overview}{a, b}). Designed to be highly flexible and adaptable, RRP supports daily research activities and accommodates the often non-linear progression of projects. It enables continuous tracking and shareability, with all work logged and constantly accessible, and supports reproducible execution across cloud, institutional, local, settings with consistent provenance.\\
It is specifically designed to also work with large datasets managed through a research data management system (RDMS), allowing data to be mounted and accessed directly (\figref{fig.rrp_overview}{b}). As RDMS, we utilize openBIS, a modular, open-source platform established in many academic institutions that combines FAIR data management with electronic laboratory notebook (ELN) and laboratory information management system (LIMS) functionality to support the documentation, organization, and sharing of scientific data, protocols, and materials \autocite{barillari_openbis_2016}. In principle, any other RDMS system could be used, provided it allows for data access by a well-defined, open protocol. However, we use openBIS due to its proven performance across dataset types and sizes, including even extremely large and complex data, namely proteomics or imaging \autocite{bauch_openbis_2011, cuny_high-resolution_2022, cuny_live_2022}. RRP projects are isolated within dedicated instances, typically managed at the research group level, inheriting the RDMS’s access controls to ensure secure, auditable, and resource-efficient operation.\\ 
For code provenance and version control, we leverage Git as the foundation of an RRP project. Code, ranging from data processing scripts and analysis workflows to full software libraries, can be integrated into projects to accommodate complex dependencies or active software development via Git submodule linking, all without disrupting existing development workflows. By building on established open-source technologies that minimize interference with user habits, RRP enables seamless integration and supports widely used graphical user interfaces (GUI), like R-Studio and VS Code.\\
Once project data is registered in the RDMS and code in Git, RRP’s integrated structure and multi-functionality enable diverse research workflows, including project and lab management, collaborative analysis, manuscript preparation, and teaching, through modular integrations and shared environments (\figref{fig.rrp_overview}{b}). This approach lowers barriers for users spanning the technical spectrum while embedding executable environments within a unified data and software management platform. As a result, RRP advances FAIR-compliant research practices and reproducibility in contemporary science.\\
Although best deployed on group servers, RRP runs on all major operating systems (\textbf{Methods}). Crucially, one-click launch manifests enable users to reproduce or extend published work from where previous efforts left off without needing an RRP server installation, accelerating reproducibility, future research, and promoting knowledge, data, and code reuse. Taken together, we structured RRP around projects, enabling researchers to manage all key aspects of a research project within a single interface, ensuring isolation, resource efficiency, and a robust, yet flexible environment for modern, by design FAIR-compliant research practices.

\subsection{RRP system architecture}

\begin{figure*}[ht!]
    \centering
    \includegraphics[width=16cm ]{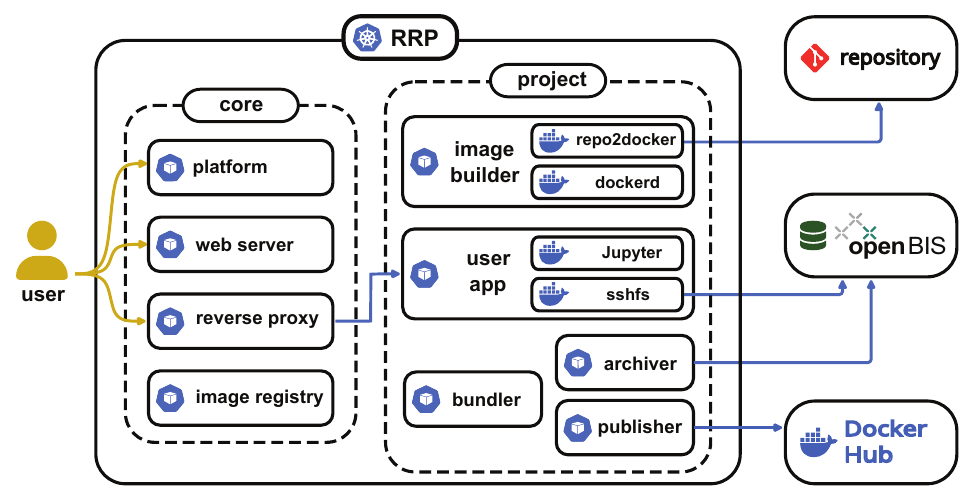}
    \caption{Schematic of RRP's system architecture. RRP's components are grouped into ``core'' and ``project'' and interactions with RRP project specifications (Git repository), RDMS (openBIS), and public Open Container Initiative (OCI) registry (Docker Hub) are indicated with arrows.}
    \label{fig.architecture}
\end{figure*}

RRP is a web application designed to run on a Kubernetes cluster with attached persistent storage. Its microservice-like components are organized as loosely coupled Kubernetes pods, each comprising one or more Docker/Open Container Initiative (OCI) containers. For an RRP instance, all components deploy via declarative Argo CD configuration. Components are grouped into the RRP ``core'' and those handling user ``projects'' (\figref{fig.architecture}). Multiple user projects can run concurrently, sharing compute resources that Kubernetes scales elastically. Among the core components, the ``platform'' serves the RRP back-end, while an Apache ``web server'' serves the front-end. Front- and back-end communication is done through an internal REST API and a server-sent event stream. A ``reverse proxy'' redirects incoming web connections to the appropriate user project front-end. The ``image registry'' serves container images built for user projects.\\
The RRP platform builds projects from a public or private Git repository URL. Alongside the project-specific code and required data, the repository must include a definition of the computational environment per Reproducible Execution Environment Specification (REES) \autocite{meng_facilitating_2017, huch_reusable_2025}. REES is a formal specification to define the elements necessary to recreate the exact software environment used in a scientific workflow. The ``image builder'' component runs repo2docker \autocite{jupyter_binder_2018} to build a container image from the user-provided RRP Git repository. The ``user app'' is that container deployed as an independent pod to the Kubernetes worker nodes. It connects to an external RDMS instance (openBIS) to mount RRP project datasets directly via its built-in secure file transfer protocol (SFTP) server. It saves its state, including a working copy of the analysis code originally found in the repository, in a project-specific folder on the persistent file storage. RRP includes three key components for preserving and sharing reproducible research projects. The ``archiver'' component can, on demand, store results or the entire project, including data and the computational environment via its container image, in the attached openBIS instance for future retrieval. The ``publisher'' can push the container image to a public OCI container registry, including Docker Hub. The ``bundler'' exports a complete, standalone archive, called a ``player bundle'', essentially a compressed file containing all input data, the container image, and the analysis code. This archive enables full offline reproducibility; recipients with Docker can use the included platform-specific startup script to recreate the entire computational environment and datasets locally, even without a complete RRP installation. It can be shared with anyone within or outside the research group (e.g., reviewers, colleagues, the public), allowing them to reproduce the complete analysis environment and reuse the research work. Even easier to share is the lightweight ``player script``, a minimal bundle containing only the startup script and metadata. It reproduces the computational environment locally by downloading data and code artifacts from online sources (e.g., Zenodo) that have already been published with a persistent, public digital object identifier (DOI), using a small launch script. Depending on dataset size, data can be bundled directly (for small datasets, a few GB), retrieved from public repositories (for datasets, a few 100 GB), or mounted live from RDMS storage (for large datasets, assuming a robust network connection). In all cases, users interact with RRP projects using their local browser. This modular, containerized architecture ensures scalable, reliable, and reproducible environments, seamlessly integrating data and code, ultimately enabling flexible, robust research practices across diverse computational infrastructures.

\subsection{Stepwise building of reproducible research environments within RRP projects}

\begin{figure}[H]
	\centering
      \includegraphics[width=\textwidth]{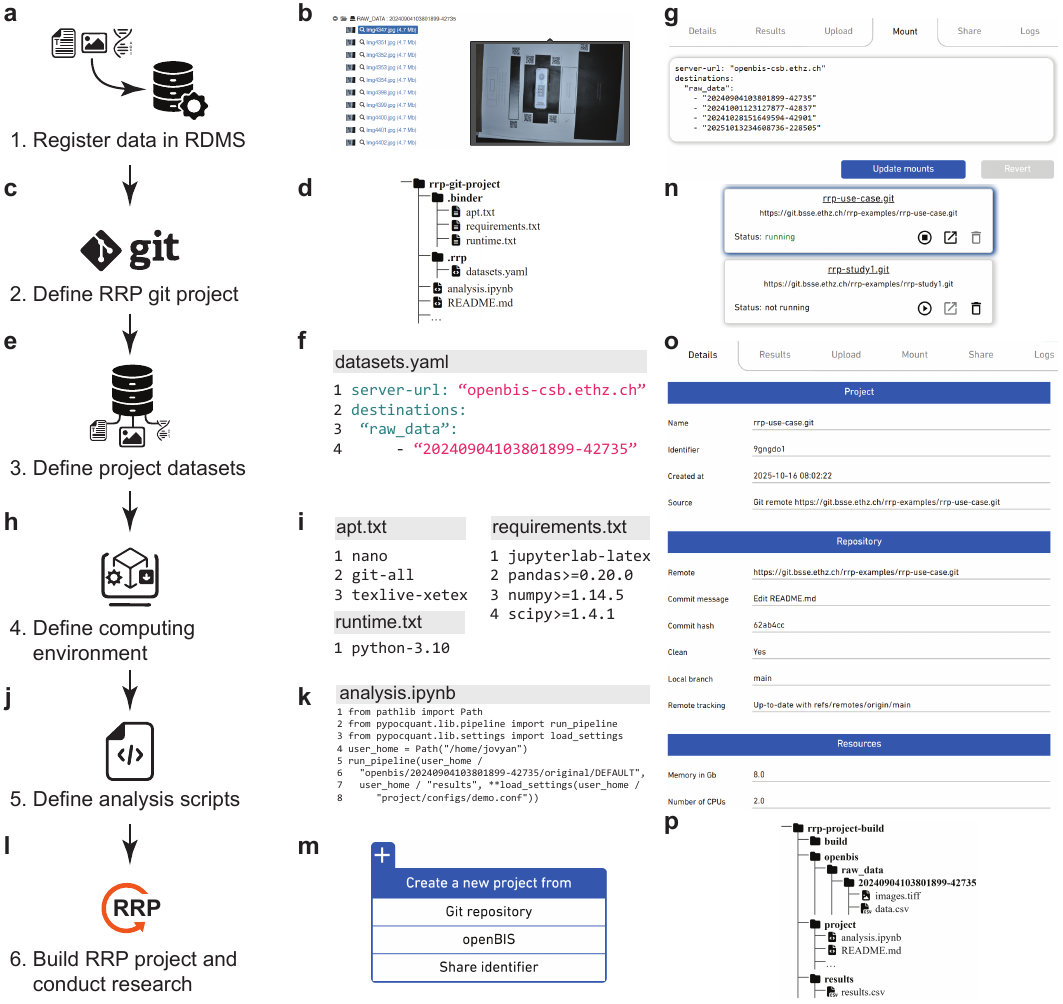}
	\caption{New RRP project specification. \textbf{a,b}, Step 1: Registering required research data in RDMS generates unique identifies (permID) for the datasets. \textbf{c,d} Step 2: Definition of a RRP Git project. The example folder listing shows the minimally required folders and optional files in the repository root. \textbf{e,f} Step 3: Definition of datasets to be mounted in RRP form the RDMS. \textbf{f} Example of a datasets.yaml file with the server, the folder name of the dataset on RRMS, and the permID of the dataset. \textbf{g} New data can be added any time, for example, in the RRP GUI. \textbf{h,i} Step 4: Definition of the computation environment for the project. \textbf{i} Example of system libraries (apt.txt), runtime version (runtime.txt), and packages (requirements.txt). \textbf{j,k} Step 5: Optionally, an analysis notebook/workflow (\textbf{k}) describing how data is transformed into results, as well as additional files relevant to the research project, can be added to be tracked. \textbf{l,m} Step 6: With an RRP Git project fully defined, it can be created (built) from the RRP GUI (\textbf{m}). It is also possible to open an existing RRP project present in the RDMS (openBIS) or form a shared identifier, when shared from a colleague. \textbf{n} The project's computational environment is then build and the project can be started, stopped or deleted. \textbf{o} In the Details Tab of the project one gets an overview over the Git status and can allocate the required computational resources (CPU, RAM). \textbf{p} When entering the project, the Git repository from \textbf{d} is now built into a container and its file contents are present in the /project subfolder. The data is mounted into the /openbis subfolder and results can be saved in the /results subfolder.}
	\label{fig.rrp_new_project}
\end{figure}

A typical workflow begins with (1) registering research data and metadata in the RDMS, if they already exist at this research stage, e.g., from earlier experiments, collaborators, or reuse (\figref{fig.rrp_new_project}{a,b}).
To ensure provenance and facilitate collaboration, an RRP project is version-controlled and typically defined as (2) a Git project hosted on a public or institutional GitHub or GitLab instance (\figref{fig.rrp_new_project}{c}). If the repository is private, access credentials must be provided through the RRP GUI. Users are flexible in structuring the Git repository; however, a minimal RRP project Git repository includes the \textit{.binder} folder specifying the computing environment per REES and the \textit{.rrp} folder containing the required datasets definition (\figref{fig.rrp_new_project}{d}, \tableref{table.rrp_source_table}). \\
To include data (3), users specify the RDMS server URL, dataset permIDs, and target locations within the RRP project (e.g.,  raw\_data folder) in \textit{dataset.yaml} (\figref{fig.rrp_new_project}{e,f}). RRP projects scale straightforwardly to large datasets (hundreds of GBs to TBs) because they mount data directly from the RDMS as a data storage backend. This version-controlled file can evolve with the project and may initially be empty, but revised anytime in the RRP Git project or GUI to mount additional data (\figref{fig.rrp_new_project}{f,g}). Alternatively, small to medium-sized datasets can be stored directly in the project's Git repository, e.g., via Git LFS. \\
Next, we define (4) the computing environment, including customization of the RRP user experience (R-Studio, VS Code, ...) and third-party libraries, run-time dependencies, and language kernels (Python, R, Julia, C++, Matlab, ...) via ecosystem-specific package managers (Pip, Conda, CRAN, Apt, Nix) (\figref{fig.rrp_new_project}{h,i}). To do so, RRP requires a set of REES-compliant configuration files to define the computation environment stored in \textit{.binder} (\figref{fig.rrp_new_project}{i}).
Examples include \textit{requirements.txt} (\textit{Pip}) for customizing the user experience, \textit{Project.toml} (Julia), and \textit{apt.txt} (\textit{Apt}) for software packages. System tools like Git, LaTeX, or Nano are added via \textit{apt.txt}, and the interpreter version (e.g., Python 3.10) is specified in \textit{runtime.txt}. Additional language-specific files, namely \textit{install.R}, define R packages. \\
Analysis code (5) can be organized freely within the repository. If present at project creation, analysis scripts or notebooks can be added (\figref{fig.rrp_new_project}{j,k}). To ease RRP Git project creation, we provide cookie-cutter templates for popular languages ( \tableref{table.rrp_projects_table} or the RRP documentation).\\ 
At this point, an RRP Git project is defined, and RRP can be accessed in a browser. Users can then log in to RRP using the RDMS credentials to (6) create the project (\figref{fig.rrp_new_project}{l}, \extref{fig.rrp_si_fig3}[a-d]). New projects can be created from scratch using an RRP Git repository or existing ones added from openBIS or using a shared identifier (\figref{fig.rrp_new_project}{m}). Once the RRP platform clones the RRP Git project, RRP automatically builds the computational environment by mounting required datasets (\figref{fig.rrp_new_project}{f,g}) and generates a Docker image encapsulating all dependencies, tools, research data, and code (\figref{fig.rrp_new_project}{f,g,i,k}). Once the RRP platform has built a project's Docker image, it can be launched by clicking the play icon or deleted via the trash icon (\figref{fig.rrp_new_project}{n}). Each project exposes six tabs in the RRP main UI: Details, Results, Upload, Mount, Share, and Logs (\figref{fig.rrp_new_project}{o}, \extref{fig.rrp_si_fig3}[c]). The Details tab displays the project overview, repository metadata, and allows users to specify necessary compute resources (e.g., RAM and CPUs) for each project (\figref{fig.rrp_new_project}{o}). RRP is compatible with high-performance computing clusters for compute-intensive projects. A running project is indicated by its status, and clicking the tab icon opens a new session to work in the user-defined environment (\figref{fig.rrp_new_project}{n}). By default, a browser-based JupyterLab instance launches, enabling interactive or command-line-based analyses with all the necessary data and code preloaded (project folder structure after the build, \figref{fig.rrp_new_project}{p}, \extref{fig.rrp_si_fig3}[d]). 

However, RRP allows users to customize JupyterLab to launch alternative popular environments, like RStudio, VS-Code, or add support for domain-specific tools. The possibilities are nearly limitless and adaptable to the research needs (see examples and \textbf{Extended Data Figures}).\\

\subsection{One‑step collaboration and sustainable reproducible research enabled by the RRP platform}

\begin{figure}[H]
	\centering
      \includegraphics[width=16cm ]{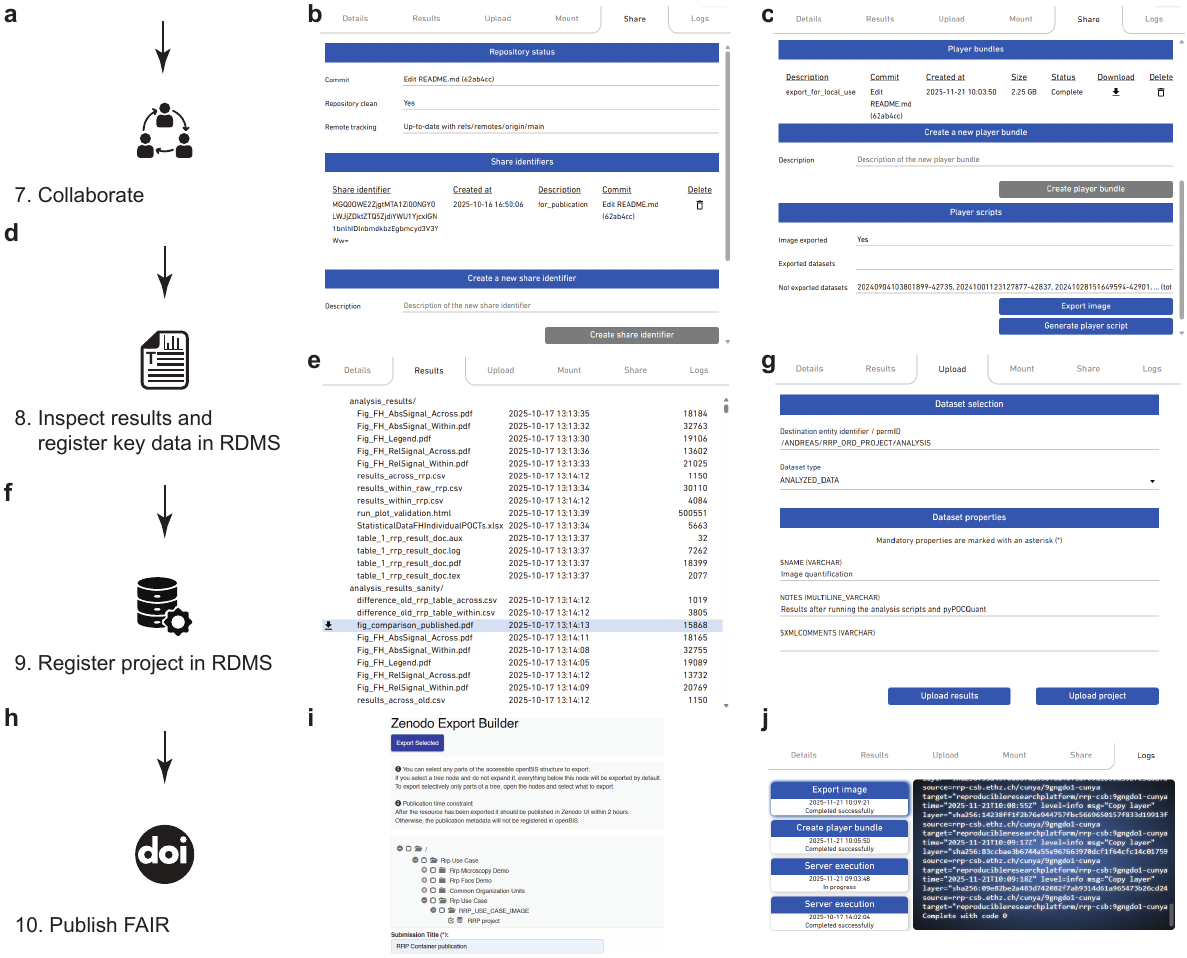}
	\caption{Steps for collaboration and publishing an RRP project. \textbf{a,b}, For collaboration (optional Step 7), one can share the current stage of the project by generating a Share identifier from within the Share tab of an RRP project within the GUI. Others do not have to build the computational environment and simply obtain a clone. \textbf{c} To share the RRP project with anyone or for offline use (e.g., outside the research group), one can bundle the Project into a player bundle (that includes all the data) or generate a Player script (without data, reduced archive size). However, the latter requires exporting the datasets first. Finally, the whole image can be exported to the RDMS to snapshot an important milestone. \textbf{d,e} To inspect the results of an RRP project (optional Step 8), one does not need to start it. When saved within the /results folder, they can be inspected from the RRP GUI. \textbf{f,g} Registering an RRP project or any results in the RDMS can be done form the Upload Tab in optional Step 9. \textbf{g} Here, any data-type can be registered in the RDMS and the RRP project be attached to a project in the ELN-LIMS. \textbf{h,i} In Step 10 the RRP project can be published to obtain a DOI (e.g. Zenodo) through the RDMS. \textbf{j} The Logs tab lists information from the different interactions with an RRP project.}
	\label{fig.rrp_ui_implementation}
\end{figure}

RRP is designed for collaboration (7). RRP projects can be shared at any point via a shareID, instantly making them available to users on the same RRP instance, without requiring the project to be rebuilt (\figref{fig.rrp_ui_implementation}{a,b}). To ensure consistency and reproducibility, the RRP Git repository must be clean (i.e., all local changes committed) before sharing and exporting (\figref{fig.rrp_ui_implementation}{b}). RRP supports exporting projects via either “player bundles” or “player scripts” (\textbf{RRP System Architecture}). Users select the appropriate export based on dataset size and collaboration context: player bundles are full, self-contained, executable, offline-capable containers including all data, code, and a working computational environment, while player scripts rely on online data availability for more lightweight sharing (\figref{fig.rrp_ui_implementation}{c}). Upon launching a player script on a local machine, RRP downloads necessary data from remote repositories, unpacks resources, and opens the computational environment in a local browser. This allows anyone to reproduce or interact with the research projects seamlessly without rebuilding environments or installing additional software.
Essential research project outputs (e.g., results, figures, tables) can (8) be saved in a special subfolder (\textit{results}) and appear in the Results tab and can be inspected or downloaded without launching the project (\figref{fig.rrp_ui_implementation}{d,e}). These project outputs can be registered with openBIS to obtain a permID and attach appropriate metadata to ensure long-term preservation directly via the RRP user interface (\figref{fig.rrp_ui_implementation}{g}). As projects evolve, data and metadata can be updated, and changes to scripts or notebooks are tracked in the RRP project through Git, ensuring versioning and provenance. Within the RRP project session (e.g., JupyterLab), users can also interact with the RDMS to register data or obtain metadata through tools namely the Python API pyBIS\autocite{openbis-team_python_2025}. \\
Upon reaching a research project milestone or finalization, the state of the RRP project can (9) be made persistent by associating it with an ELN entry and archived to the RDMS (\figref{fig.rrp_ui_implementation}{g}). Therefore, the entire executable container, including the computational environment, data, and code, is persistently saved.\\
Lastly, (10) a permanent DOI can be issued via the RDMS from Zenodo or institutional repositories to make an RRP project permanently publicly available (\figref{fig.rrp_ui_implementation}{h,i}). The exported RRP project is fully FAIR-compliant and executable by anyone, ensuring the work's transparency, reproducibility, and reusability, including data and analysis.\\
The Logs tab in the RRP GUI helps diagnose issues, e.g., when using obscure or outdated packages within the RRP Git REES, and to monitor the project cloning, building, or execution process (\figref{fig.rrp_ui_implementation}{j}). \\
Taken together, RRP reduces the complexity of reproducible research to just a few intuitive steps by automating non‑trivial setup tasks, defining consistent computational environments, and seamlessly integrating big data directly from an RDMS without downloads or risk of local data loss. RRP projects remain flexible to evolving definitions and data, while enabling real‑time collaboration in shared environments and ensuring direct reuse of work in future.\\

\begin{figure}[H]
	\centering
      \includegraphics[width=16cm ]{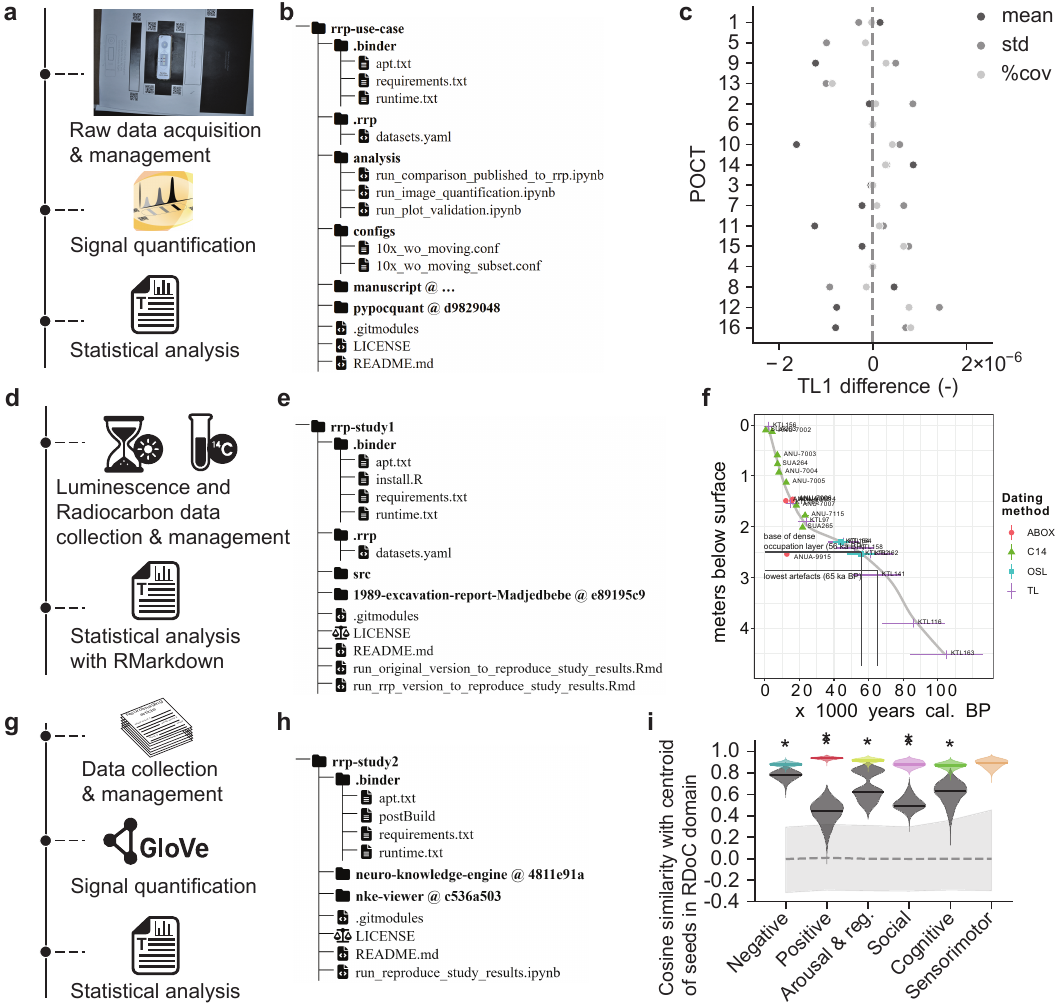}
	\caption{Examples of reproducing published research within RRP. \textbf{a}, Workflow of the research from raw data management to statistical analysis in quantifying rapid diagnostic test line signals. \textbf{b}, RRP Git folder content. \textbf{c},  Differences of the result obtained within RRP to the published results\autocite{cuny_pypocquant_2021}. The mean  (dark grey), SD (grey), and percent coefficient of variation (light grey) are plotted for each POCT and TL1. \textbf{d} Workflow of research from data collection to analysis in the field of Archaeology. \textbf{e}, RRP Git folder content with two analysis notebooks for directly reproducing the published study and one with the data registered to our RDMS. \textbf{f} Resulting figure generated in RRP corresponding to Fig. 5 in \autocite{clarkson_archaeology_2015}. \textbf{g} Workflow of research from data collection to analysis in the field of Neuroscience and ML. \textbf{h} RRP Git folder content mounting published work. \textbf{i} Resulting figure generated in RRP (with adjusted size) corresponding to Fig. 3 \autocite{beam_data-driven_2021}. }
	\label{fig.rrp_use_case}
\end{figure}

\subsection{RRP accelerates reproduction of published work across disciplines and computational reporting practices}

To demonstrate RRP’s utility, we reproduced the results of published work that includes an image data set, a custom-developed scientific software package, and analysis code (\figref{fig.rrp_use_case}). 
This study quantified signals in images of point-of-care COVID-19 tests (POCT) using a Python tool, pyPOCQuant \autocite{cuny_pypocquant_2021}. The main data of this study were images of POCTs that were acquired and analyzed in a validation experiment. To reproduce the results, the workflow includes the raw data (images), which need to be processed with the pyPOCQuant software to quantify the signals of the test line (TL1) and the control line (CTL), and then their statistical analysis (\figref{fig.rrp_use_case}{a}). To reproduce the results in RRP, we first registered the image data in an openBIS instance; then we created a new RRP Git project and defined the computational environment, including the necessary dependencies (\figref{fig.rrp_use_case}{b}). Besides the mandatory \textit{.rrp}, \textit{.binder} folders and configuration files, the project includes an \textit{analysis} folder with two Jupyter notebooks: \textit{run\_image\_quantification.ipynb} to quantify the signals from the raw data and \textit{run\_plot\_validation.ipynb} to perform the statistical analysis. The required software pyPOCQuant is added as a submodule to the RRP Git project to be used directly from the source code. Running \textit{run\_comparison\_published\_to\_rrp.ipynb} allowed us to compare the reproduced results to the original publication (\figref{fig.rrp_use_case}{c}, \extref{fig.rrp_si_fig5}[a,b]). Apart from the expected floating point differences in the range of $10^{-4}$ to $10^{-6}$ caused by the different underlying hardware, we reproduce the results in RRP successfully (\figref{fig.rrp_use_case}{d}).
We further reproduced two additional studies: a decade-old archaeology analysis using R \autocite{clarkson_archaeology_2015} (\figref{fig.rrp_use_case}{d-f}) and a recent neuroscience and machine learning study relying on deprecated Python 3.6 \autocite{beam_data-driven_2021} (\figref{fig.rrp_use_case}{g-i}). First, we achieved direct reproduction by adding the published code and data as a git submodule and running \textit{run\_original\_version\_to\_reproduceduce\_study\_results.Rmd} in R-Studio (\figref{fig.rrp_use_case}{e–f}). Second, we registered the data in our RDMS and modified the code so that RDMS-mounted data were used, with results saved to the RRP 'results' folder.
The second example highlights how RRP enables reproduction of studies conducted with outdated package versions and Python, illustrating both the ability to reproduce and to extend prior work (\figref{fig.rrp_use_case}{h–i}). Overall, we demonstrate that published studies from diverse domains and time periods can be translated to RRP with minimal effort; once data are registered in the RDMS, results become FAIR, readily reusable, and indistinguishable from work initially conducted. Thus, future users are freed from the burden of configuring legacy computational environments.

\subsection{RRP unifies open science tools to manage scientific projects across disciplines and throughout the research lifecycle}

To demonstrate RRP's potential as a single platform for managing scientific projects and collaboration, including publication, we include this manuscript in our RRP use-case project and additionally show the round-trip with a minimal example (\extref{fig.rrp_si_latex}). We created a new project on Overleaf, an open-source, real-time collaborative LaTeX editor with Git integration, to write the manuscript \autocite{the_overleaf_team_open-source_2014}. We then exported this project to GitHub and added it as a Git submodule to the RRP repository. This setup enabled manuscript editing directly within the RRP environment using JupyterLab, without leaving the platform. 
LaTeX was added to \textit{apt.txt} and jupyterlab-latex to \textit{requirements.txt} to enable an in-environment editor and Zotero integration managed references. This direct link with Overleaf, publishers, and archives significantly facilitates and accelerates the publishing process (\extref{fig.rrp_si_latex}[e]). It eliminates the need to manually copy manuscript files for publication, keeping them alongside data and code, turning the process into a one-click procedure. For the associated data, analysis, and computational environment, a DOI can be obtained by publishing the RRP 'player bundle' to a data repository like Zenodo (\extref{fig.rrp_si_latex}[f]). \\
RRP, as a single platform, unlocks the potential of open science across disciplines. Because openBIS ELN-LIMS can manage and organize data from any quantitative discipline, RRP can be used to process such data in a reproducible and open way, building on other open science tools. For example, we use QUEEN, a reproducible DNA editing tool for life sciences\autocite{mori_framework_2022}, to simulate the steps of a CRISPR protocol \textit{in silico} tagging the target protein Myo1 in yeast (\textit{Saccaromyces cerevisiae}) with a fluorescent label (3x mKate2) (\extref{fig.rrp_si_genetic_engineering}) storing all DNA materials in the RDMS. In engineering and related disciplines, RRP reproducibly supports collaborative tasks within a single environment such as computer-aided design-based hardware design, 3D prototyping, and designing electronic circuits to generate complete printed circuit board (PCB) fabrication files (\extref{fig.rrp_si_engineering}[a,b]). The RDMS helps to organize parts libraries and components and links them to projects and experimental results.
RRP also supports AI applications, including machine learning. In a sample RRP project, microscopy data from the RDMS are mounted, and a PyTorch-based library (qute) trains a custom deep-learning model, monitored with TensorBoard inside RRP, with model predictions visualized in a Jupyter notebook (\extref{fig.rrp_si_dl}[a,b]). 
Users unfamiliar with scripting can use large language models within RRP for assistance. For example, when asking how to count cells in an image using the previously trained model (\extref{fig.rrp_si_dl}[c]), the suggested code correctly returned a cell count of 47 (\extref{fig.rrp_si_dl}[d]).\\
RRP also offers new opportunities for teaching by providing students with immediate access to course materials and a ready-to-use computing environment, eliminating the need for time-consuming software installations and supporting isolated exercise or exam sessions. Leveraging JupyterLab’s extensive ecosystem, the platform can be enhanced with tools for drawing, visualization, presentations, and social interaction. It also integrates domain-specific tools, like Geographic Information System (GIS) editors for geosciences or molecular structure viewers for biology.\\ 
However, RRP is not limited to the JupyterLab ecosystem and allows the use of an entire desktop environment for complex workflows, e.g., including large image data that requires interaction with GUIs (\extref{fig.rrp_si_desktop}[a-d]). Using the popular RStudio, we demonstrate its use within RRP and subsequently reproduce the results within a shared ''player bundle'' executed on a local laptop running a different host operating system (macOS) (\extref{fig.rrp_si_r_usecase}[a-d]). Lastly, the RRP platform can be used to develop and maintain software using the popular VS Code IDE and its debugger (\figref{fig.rrp_si_vscode}{a,b}). Integrating openBIS ELN-LIMS as the RDMS for the comprehensive handling of data artefacts of any kind, RRP fully satisfies the FAIR principles for RDM (\extref{fig.rrp_si_mic_facs}[a-d]). The modular RRP platform provides an efficient means to work on every stage of a complex scientific project, from planning through publication. It preserves the entire working environment so that anyone can later resume the work exactly where it was left off, as though logging into the original contributor's computer.

\section{Discussion}

We demonstrate that the Reproducible Research Platform (RRP) effectively addresses key challenges in research reproducibility by bridging the separation of data and code life cycles, providing integrated executable computational environments, and lowering barriers to adoption across diverse technical skill levels.
We outline the key steps in generating and organizing a new RRP project, including project data and interaction with the RRP GUI, as well as registering newly generated results in the RDMS and utilizing collaborative functionalities to share an RRP project with collaborators and publish it for long-term preservation. 
We further demonstrate that RRP can reproduce results from three published research projects spanning over ten years, including those developed with older R and Python versions \autocite{clarkson_archaeology_2015, beam_data-driven_2021}. We demonstrate how to directly utilize published code and data, and create an RRP project or take the minimal steps to register data in the RDMS for proper annotation, future reuse, and to maintain a minimal size for the Git repository. This highlights RRP’s robustness in sustaining reproducibility across software and project evolution and easy adoption of the platform. 
Interdisciplinary projects that rely on complex systems and vast data volumes place stringent demands on RDM and platforms \autocite{hart_ten_2016, sarkans_rembi_2021, mani_genomics_2025}. Yet, many current practices and tools rely on fragmented, manual workflows \autocite{zhang_best_2022, boehm_quarep-limi_2021, montero_llopis_best_2021, gruning_practical_2018} that are difficult to scale or reproduce and often misalign with users’ habitual working patterns. 
RRP addresses these challenges through an open-source modular infrastructure that simplifies collaborative work and integrates means of persistent publication. Projects can be shared easily with colleagues or supervisors without requiring them to manually replicate data structures, adjust file paths, or install dependencies, significantly reducing overhead and fostering seamless collaboration. We hypothesize that RRP can help communities refocus on core scientific questions and accelerate knowledge generation.
Moreover, the widespread adoption of RRP has the potential to catalyze the development, evolution, and reuse of domain-specific tools, templates (e.g., machine-actionable DMPs
\autocite{jose_machine-actionable_2020}), and workflows\autocite{crusoe_methods_2022}.\\
The platform architecture scales and supports allocating computational resources (CPU, RAM) at the project level. While current capabilities may be insufficient for the largest datasets or high-end ML/AI workloads, developments to integrate RRP with high-performance computing infrastructure, including GPU clusters, are ongoing. With solutions in sight at the institutional level, reusing such large-scale projects anywhere remains a challenge, even when FAIR principles are met.
To address this, we are working toward integration with federated infrastructures namely the European Open Science Cloud (EOSC) \autocite{eosc_federation_services_2025}. These developments aim to enhance interoperability and support for additional repositories \autocite{eosc_data_commons_services_nodate}.
RRP’s technical foundation using Git to manage code and environments following REES principles, storing structured data within an RDMS, and integration with public repositories and publishing services also opens avenues for future enhancement. For example, the platform could be extended with AI-assisted interfaces, like chatbots or large language models, to simplify user interactions and automate repetitive tasks further.\\
In summary, RRP aims to make reproducible science more practical and widely accessible. It requires only a local Docker installation, sufficient disk space, and, for large datasets, a stable internet connection. With broad institutional support and appropriate computational infrastructure, this approach could significantly increase long-term reproducibility of research projects, accelerate knowledge sharing, simplify collaboration across research groups and interested parties, thereby reducing wasted efforts caused by irreproducible studies and results. 
RRP serves as a foundation for modular systems that support a FAIR scientific process and lays the groundwork for future innovations to improve usability, transparency, and reproducibility across disciplines. 

\section{Methods}

\subsection{RRP implementation}

We used a Kubernetes cluster (version 1.31.7) to deploy RRP. The RRP backend was mainly written in Scala (version 3.1.3) and some adapter code in Java (version 17). The RRP front-end is written in TypeScript (version 4.4.2), uses the React framework for the user interface components (version 17.0.1), and is served via Apache (version 2.4.52). The RRP Apache container is built on the 'ubuntu:22.04' base image (the next release will upgrade to the latest LTS version). We used RRP (version 0.5, build 21770f2) with repo2docker (version 2024.03.0) to create and run all the presented RRP projects in this study. The link to the RRP source code is listed in the Code availability section and \suppref{table.rrp_source_table}.

\subsection{RRP local deployment}

RRP can be deployed on a standard PC (Windows, macOS or Linux) for testing and evaluation purposes. To setup RRP locally, a Docker Desktop installation is required. On Windows, it was tested it with Docker Desktop 4.44.3 (202357) with WSL2 activated, and on Mac (macOS 15+) with Docker Desktop 4.48 and Docker version 28.5.1. Additionally, the setup has been tested on Ubuntu 64-bit 24.04.4 LTS running as a guest system on a Windows 11 host with Oracle VirtualBox 7.2.0-170228-Win and Oracle VirtualBox Extension Pack 7.2.0 and VBoxGuestAdditions installed. Scripts are available to install and run RRP from a Terminal or PowerShell, depending on the operating system. On Linux/macOS, shell scripts (.sh) should be run as "./scriptname.sh" after making them executable (using chmod +x scriptname.sh). On Windows, the batch files (.bat) should be run ".\textbackslash scriptname.bat" in PowerShell or Command Prompt. First, the "build" scripts should be executed. Then to start or stop RRP, the "start" script or the "stop" script. After running start, RRP is available at https://localhost/ (https://localhost/:7443 on Windows) and the RDMS openBIS at https://local.openbis.ch:8443/openbis/ with the credentials rrp-demo/rrp-demo. The RRP system can then be used for testing and demonstration purposes, for example, by running the example RRP project (\href{https://sissource.ethz.ch/sispub/reproducible-research-platform-demo-project.git}{https://sissource.ethz.ch/sispub/reproducible-research-platform-demo-project.git}), \suppref{table.rrp_source_table}. More detailed instructions for local deployment on the different operating systems and e.g. with own data from productive openBIS instances are provided in the README file of the corresponding repository (\href{https://sissource.ethz.ch/sispub/reproducible-research-platform-demo.git}{https://sissource.ethz.ch/sispub/reproducible-research-platform-demo.git}).

\subsection{Reproduction of study results}
We used image data of Covid LFAs acquired with a Nikon D90 camera mounted at a fixed distance above the LFAs from the study \autocite{cuny_pypocquant_2021}. We then registered the image data set in openBIS and mounted the data in an RRP project using their permID. We then run the Jupyter notebook on the data mounted in RRP to quantify the signals and obtain a summary CSV table `quantification\_results` and run `run\_plot\_validation` to obtain the statistics and variability of the test lines as in the original publication. To compare the results generated on RRP with the published results, we computed the difference of the mean, standard deviation (std), and percent coefficient of variation (\%cov) between the published results and the one generated on RRP using the Python notebook `run\_comparison\_published\_to\_rrp` (Python version 3.8). The additional two studies were selected unbiased from the literature, and we have no association with the authors, nor are we experts on the content. We aimed to visually compare and reproduce the main figures published in the articles. The criteria were to include an older and a more recent study, as well as different programming languages and versions. For Study 1, we wanted to test whether we were able to reproduce the results with a recent R version. We used R version 4.2.1 (2022-06-23), and the version of the original publication was R version 3.1.2 (2014-10-31). Note that within the RRP context different R versions can be tested by simply changing the version in the \textit{runtime.txt} file. All details about the environment are associated with the RRP Study 1 git project (\suppref{table.rrp_projects_table}).
For study 2, the computation environment was not defined in the git repository associated with the research article. However, in the study's Nature Research Reporting Summary, we found some information on key dependencies. Finally, we used Python 3.6.15, which matches the major release of the version mentioned by the authors (3.6.8). Further details can be found in the RRP Study 2 project (\suppref{table.rrp_projects_table}).

\subsection{RRP use-cases}
For the RRP use-cases, we generated RRP projects that included the required dependencies, such as software or extensions to Jupyter Lab, according to the respective instructions. We provide links to all the example RRP projects in the Supplementary Information (\suppref{table.rrp_projects_table}).

\section{Data availability}
The data used in this study is included in the respective RRP 'player bundles' and available from Zenodo (for DOIs see \suppref{table.rrp_projects_table}) and as data collection including code from \href{https://doi.org/10.3929/ethz-c-000787491}{https://doi.org/10.3929/ethz-c-000787491}.

\section{Code availability}
RRP was used to perform all analyses in the paper. The source code is available under \href{https://sissource.ethz.ch/sispub/reproducible-research-platform.git}{https://sissource.ethz.ch/sispub/reproducible-research-platform.git}. The instructions for local deployment and a demonstration project are listed in the Supplementary Information section \suppref{table.rrp_source_table}). The documentation can be found at \href{https://rrp.ethz.ch/}{https://rrp.ethz.ch/}. All RRP projects, including scripts for analysis, are listed in the Supplementary Information section (\suppref{table.rrp_projects_table}).

\section{Acknowledgements}
We would like to thank the Stelling group's laboratory for testing the implementation with real-world research and feedback during the development of RRP, and J. Stelling for his encouragement and support. We gratefully thank H.M. Kaltenbach for assisting with funding acquisition to finalize the development of offline sharing for RRP projects. SIS acknowledges valuable contributions to earlier prototypes of RRP of Mikolaj Rybinski, Swen Vermeul, and Gerhard Bräunlich for the development of pyBIS.

\section{Author contributions}
A.P.C. and B.R. conceptualized the project, with initial support from A.P. and F.R.. A.L. wrote the Reproducible Research Platform software and developed scripts for a local demo deployment, with help from A.P.C., H.L., J.H., M.B., and B.R.. A.P.C. created all the RRP demo projects and use cases with the help of H.L.. A.P.C., H.L., and A.L. validated the RRP software and tested RRP on various platforms. A.P.C., H.L., J.H., and A.L. wrote the user documentation. A.P.C., H.L., and J.H. wrote the manuscript. All authors read and approved the final manuscript.

\section{Funding}
The study was funded by the ETH Z\"urich. This project was also supported by the Open Research Data Programme of the ETH Domain (EPFL SCR0973431).  

\section{Conflict of Interest}
None declared.

\printbibliography

\newpage
\section{Extended Data}
\extendeddatafigures
\begin{extendeddatafiguresenv}

\begin{extendeddatafigure}[ht]
	\centering
      \includegraphics[width=16cm ]{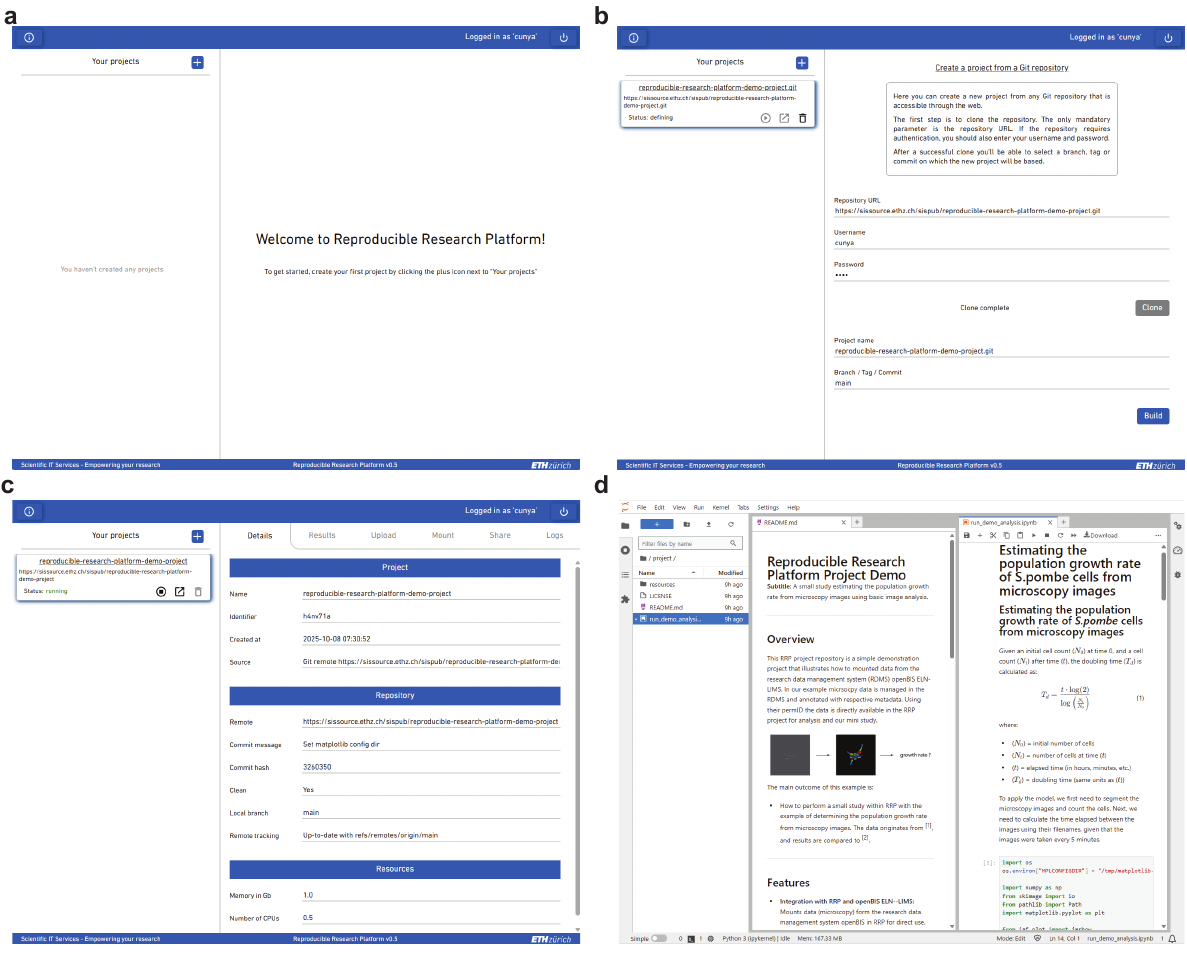}
	\caption{Creating a new RRP project with RRP's Graphical User Interface (GUI). \textbf{a}, RRP's GUI after first login, is organized with a sidebar listing all projects and a central area divided into six tabs that support key interactions with an RRP project. \textbf{b}, Creation of a new RRP project from an RRP Git repository. If the project is private, credentials are required. Users can customize the internal naming of the project and select the branch, tag, or commit of the RRP Git repository to be built. \textbf{c}, After building the project, it appears in the left sidebar. Here, the project can be started or stopped, one can enter into the computational environment (Jupyter Lab, \textbf{d}), or delete the project. Each project exposes six tabs: Details, Results, Upload, Mount, Share, and Logs. In the Details tab, the computation resources (CPU, RAM) are defined. \textbf{d}, Running the RRP demo project opens the JupyterLab interface and exposes the RRP Git repository content in the "/project" folder, where, for example, analyses can be performed, such as running the run\_demo\_analysis.ipynb Notebook.}
	\label{fig.rrp_si_fig3}
\end{extendeddatafigure}

\begin{extendeddatafigure}[ht]
	\centering
      \includegraphics[width=16cm ]{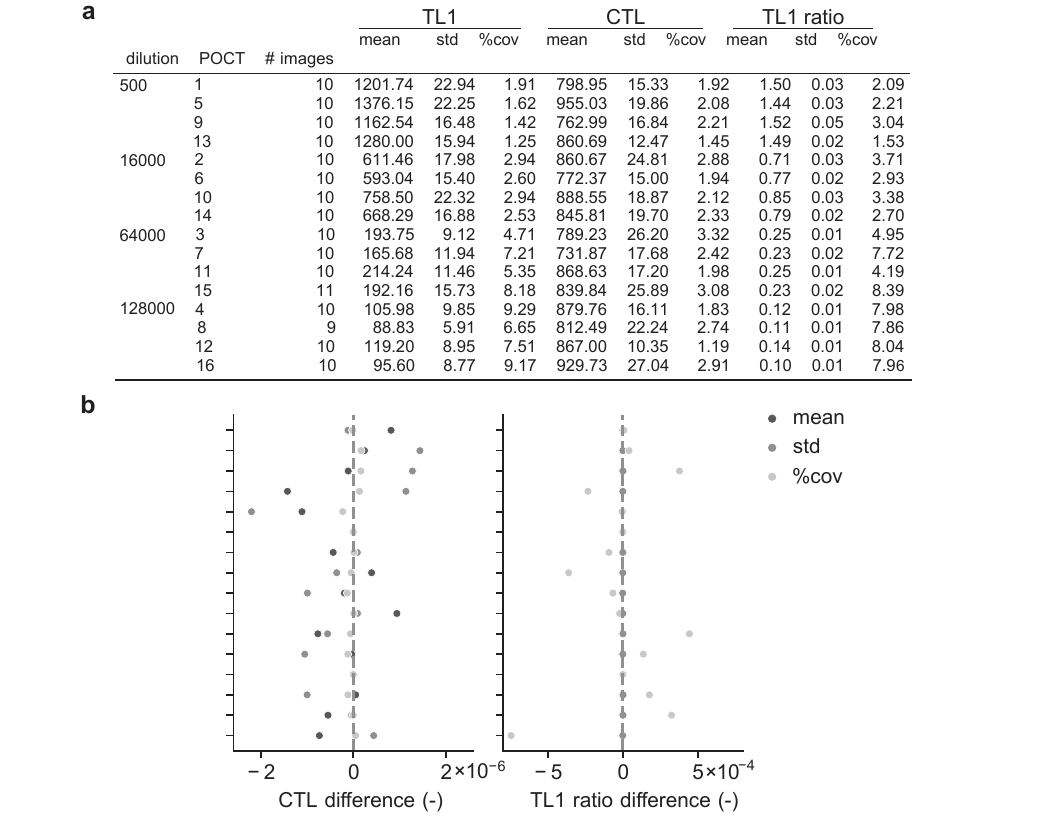}
	\caption{Reproduced results of our use-case study with RRP. \textbf{a}, Resulting table of analysis performed on RRP for the test line (TL1), the control line (CTL), and their ratio (TL1 ratio) for different dilutions and POCTs corresponding to Table S1 from \autocite{cuny_pypocquant_2021} within RRP. \textbf{b}, Differences of the result obtained within RRP (\textbf{a}) to the published results\autocite{cuny_pypocquant_2021}. The mean (dark grey), standard deviation (grey), and percent coefficient of variation (light grey) are plotted for each POCT, CTL, and TL1 ratio. The small differences can be attributed to floating-point differences and variations in third-party libraries between the Windows (study) and Linux (RRP) platforms.}
	\label{fig.rrp_si_fig5}
\end{extendeddatafigure}

\begin{extendeddatafigure}[ht]
	\centering
      \includegraphics[width=16cm ]{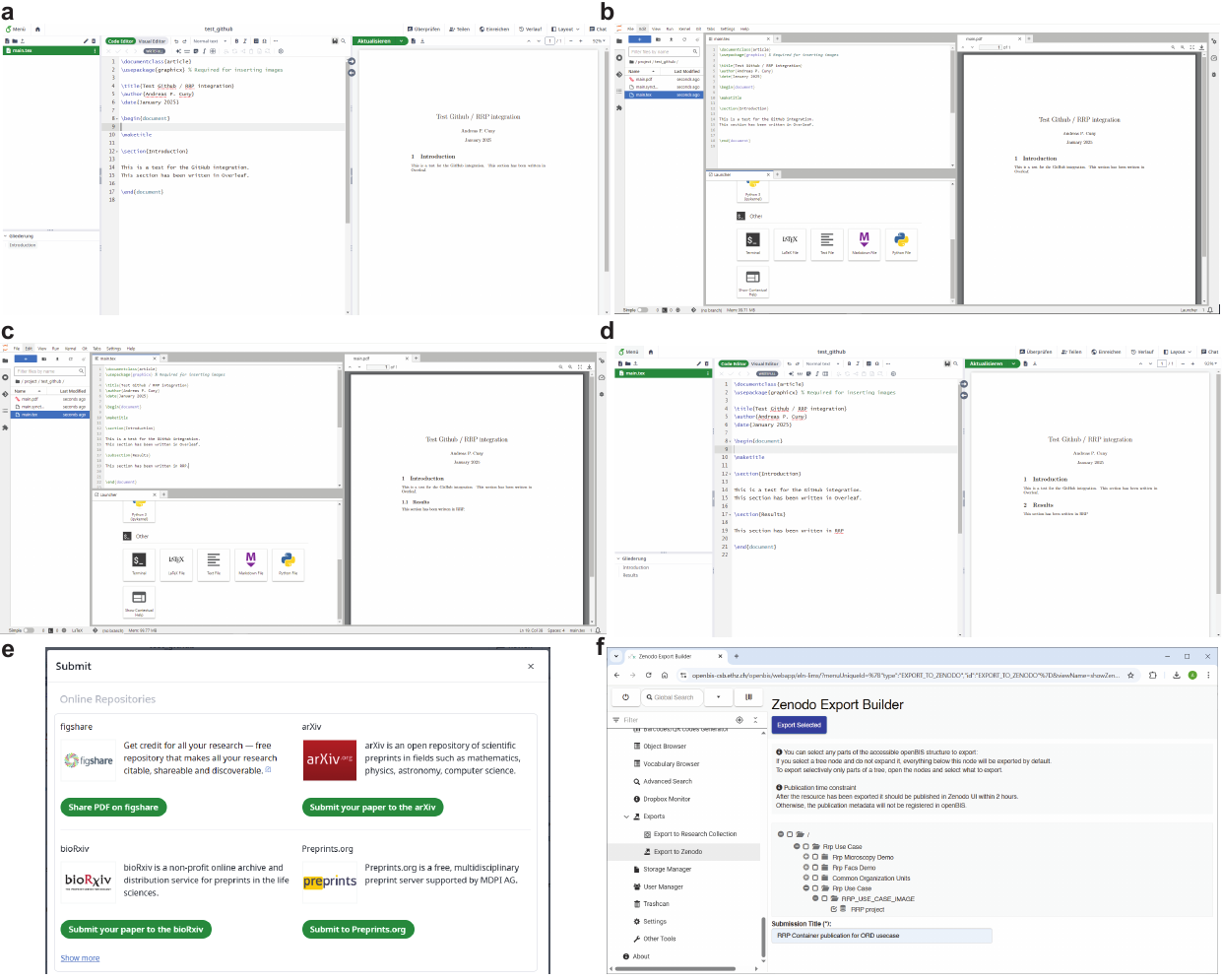}
	\caption{Example project demonstrating the use of LaTeX within RRP and integration with cloud-based services. \textbf{a}, A new LaTeX project is generated in Overleaf and then synchronized to GitHub. \textbf{b}, The Overleaf GitHub repository added as a git sub-module to an RRP project allows for direct work on LaTeX documents within the JupyterLab environment through the LaTeX extension. \textbf{c}, New content generated in RRP can be pushed back to the Overleaf GitHub project. \textbf{d}, Work added in RRP is synchronized with the Overleaf UI to finalize, e.g., a publication manuscript and push it to a pre-print server or Journal directly. \textbf{e}, Submission wizard for manuscript on Overleaf to repositories and journals. \textbf{f}, Submission wizard in openBIS ELN-LIMS for publishing an accompanying RRP 'player bundle' reproducing the results of a manuscript to e.g., Zenodo.}
	\label{fig.rrp_si_latex}
\end{extendeddatafigure}

\begin{extendeddatafigure}[ht]
	\centering
      \includegraphics[width=16cm ]{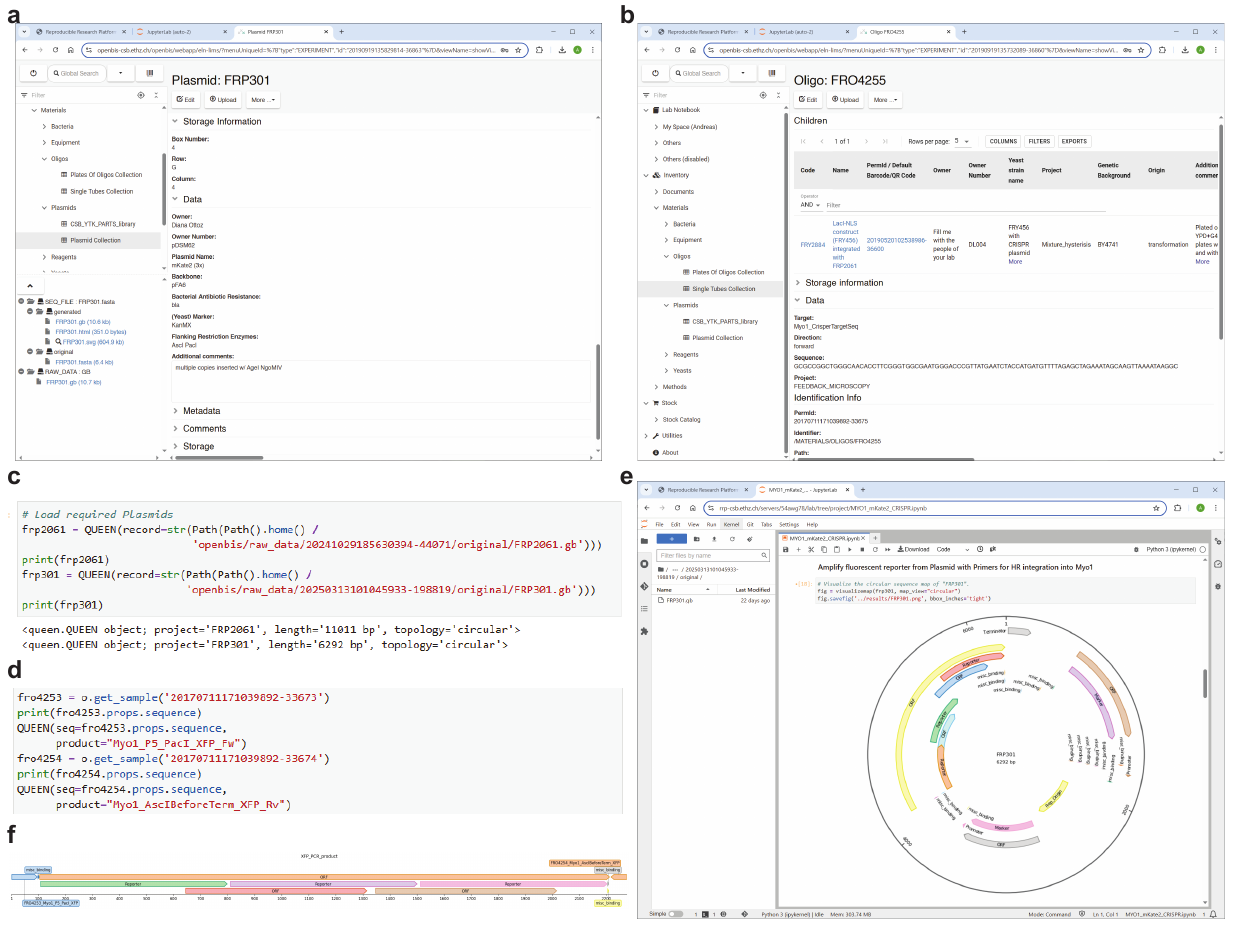}
	\caption{Example of reproducible genetic engineering with RRP and QUEEN\autocite{mori_framework_2022}, showcasing the power of using a LIMS for RDM in life sciences laboratories. This example project demonstrates the use of an in-house CRISPR protocol for C-terminal tagging of the Myo1 protein in the yeast \textit{Saccharomyces cerevisiae} with three copies of the fluorescent reporter mKate2 to engineer strains for cell cycle research \autocite{cuny_cell_2022, cuny_high-resolution_2022}. \textbf{a,b}, In the RDMS openBIS LIMS, the lab's plasmids, oligonucleotides, and chemicals are registered. \textbf{c,d}, Plasmids with annotations in the form of GeneBank files (`.gb`) can be directly mounted, while additional metadata, such as primer sequences, are loaded from openBIS. \textbf{e}, The plasmid FRP301 bearing a mKate2 fluorescent protein is shown in an RRP project. \textbf{f}, The resulting \textit{in silico} PCR product of using the plasmids and primers from \textbf{c,d}, after simulation of the CRISPR protocol.}
	\label{fig.rrp_si_genetic_engineering}
\end{extendeddatafigure}

\begin{extendeddatafigure}[ht]
	\centering
      \includegraphics[width=16cm ]{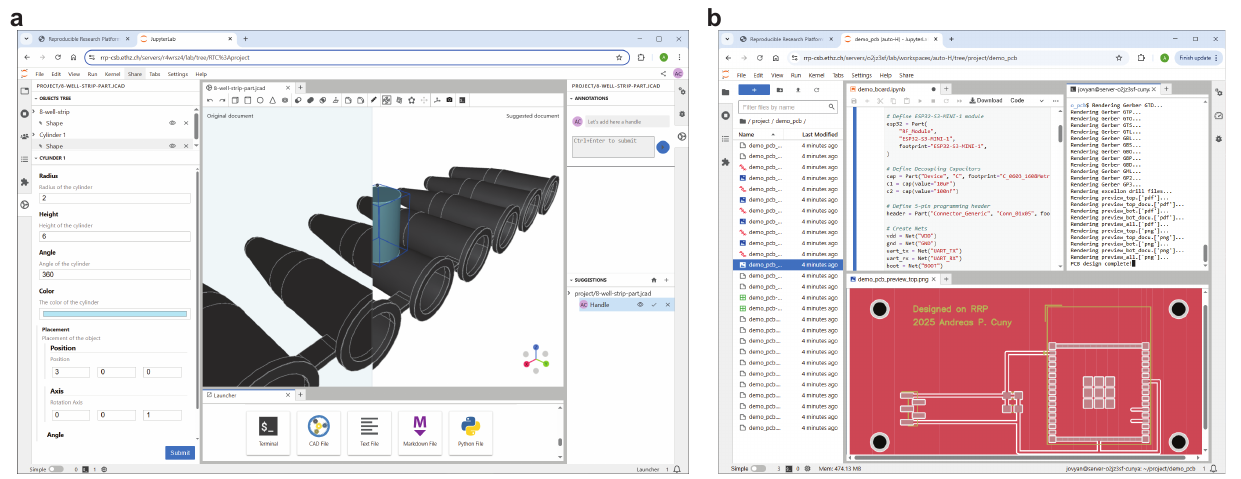}
	\caption{RRP supports the launch of full Linux Desktop environments, e.g. for interactive engineering applications. \textbf{a}, An RRP project for collaboration on CAD files. Using the desktop environment with open-source CAD drawing software such as FreeCAD or OpenSCAD allows for further flexibility. \textbf{b}, An RRP project for the reproducible design of electronic circuits with an example using an ESP32 IC. Users preferring a GUI can use KiCad within the desktop environment.}
	\label{fig.rrp_si_engineering}
\end{extendeddatafigure}

\begin{extendeddatafigure}[ht]
	\centering
      \includegraphics[width=16cm ]{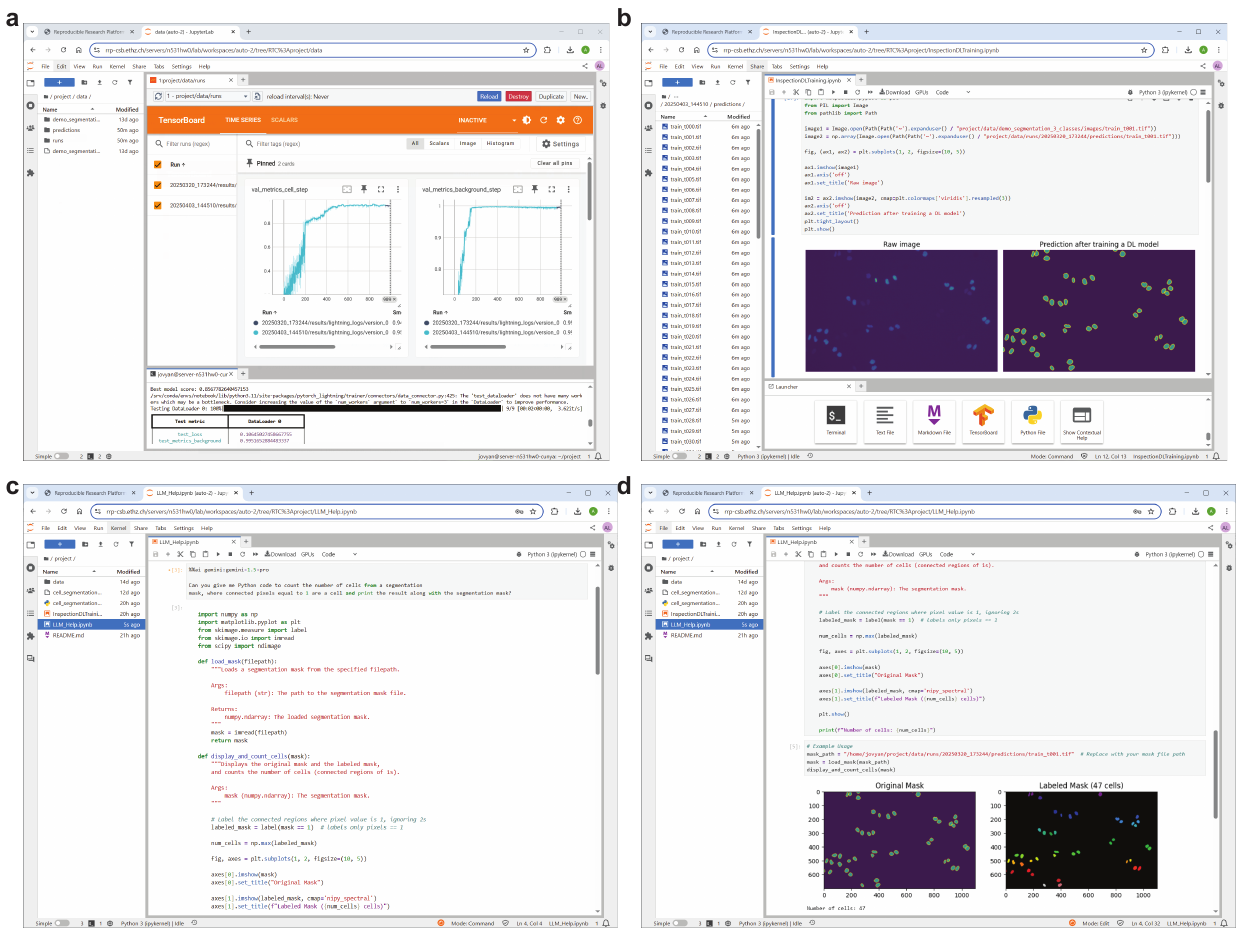}
	\caption{Example of AI on RRP, more specifically the use of deep learning for image segmentation and large language models to help with analysis. \textbf{a}, Tensorboard showing the DL model training progress for image segmentation with qute and PyTorch. \textbf{b}, Inspection of newly trained model predictions compared to input data. \textbf{c}, Example of using an LLM, e.g., assisting untrained students with image analysis, asking for Python code to count the number of segmented cells. \textbf{d}, Result of suggested code from (\textbf{c}) reports the correct number of cells when applied to the prediction of our custom-trained model.}
	\label{fig.rrp_si_dl}
\end{extendeddatafigure}

\begin{extendeddatafigure}[ht]
	\centering
      \includegraphics[width=16cm ]{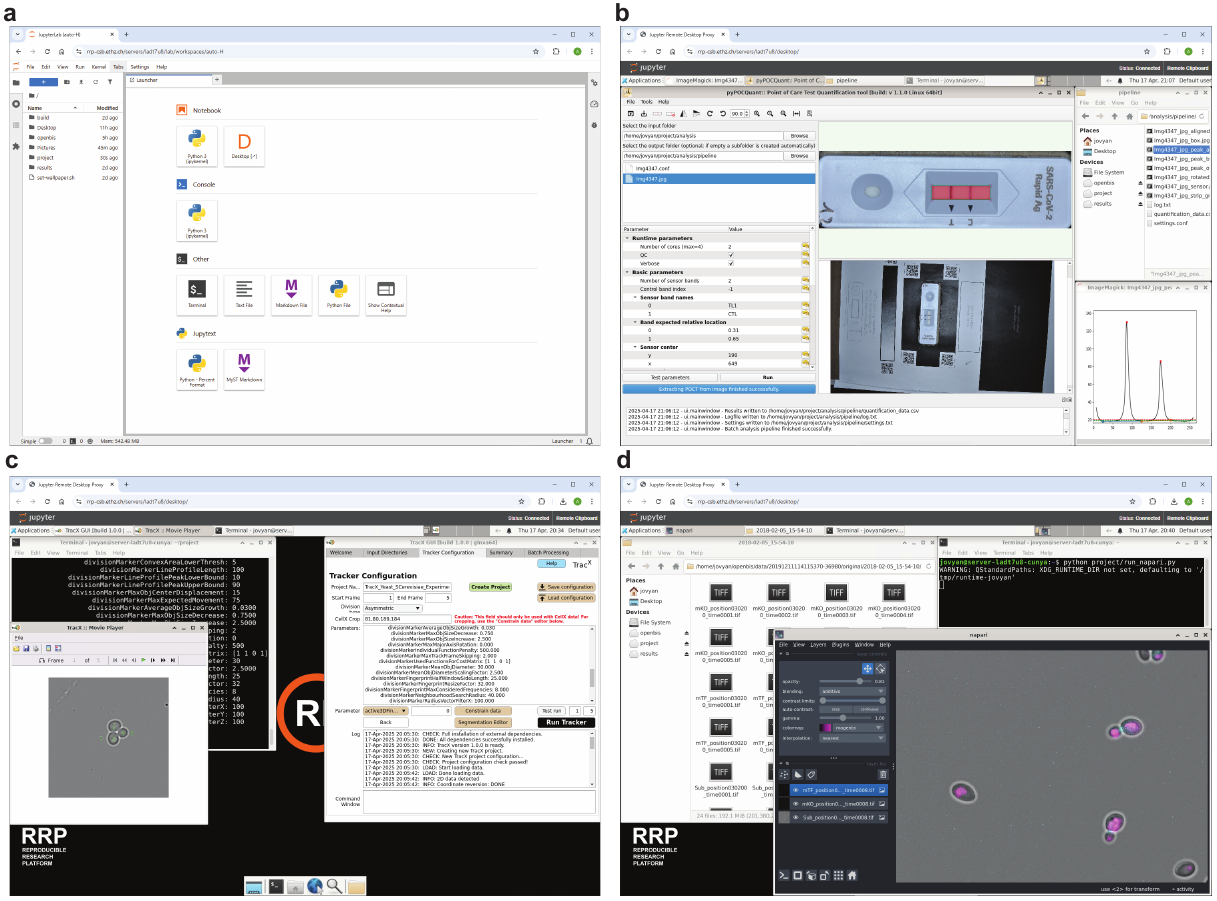}
	\caption{RRP allows to start Linux Desktop environments. \textbf{a}, RRP with the lightweight xfce4 Linux Desktop installed. \textbf{b}, On RRP, the stand-alone executable with a graphical user interface of pyPOCQuant\autocite{cuny_pypocquant_2021} is installed successfully. It allows users preferring a GUI to quantify signals of rapid diagnostic tests as presented in our use-case (\figref{fig.rrp_use_case}{}). \textbf{c}, Demonstration of the cell tracking tool $Trac^X$\autocite{cuny_cell_2022} written MATLAB run on RRP. \textbf{d}, Demonstration of using napari\autocite{sofroniew_napari_2025}, a fast, interactive viewer for multi-dimensional images in Python run on RRP interacting with microscopy data.}
	\label{fig.rrp_si_desktop}
\end{extendeddatafigure}

\begin{extendeddatafigure}[ht]
	\centering
      \includegraphics[width=16cm ]{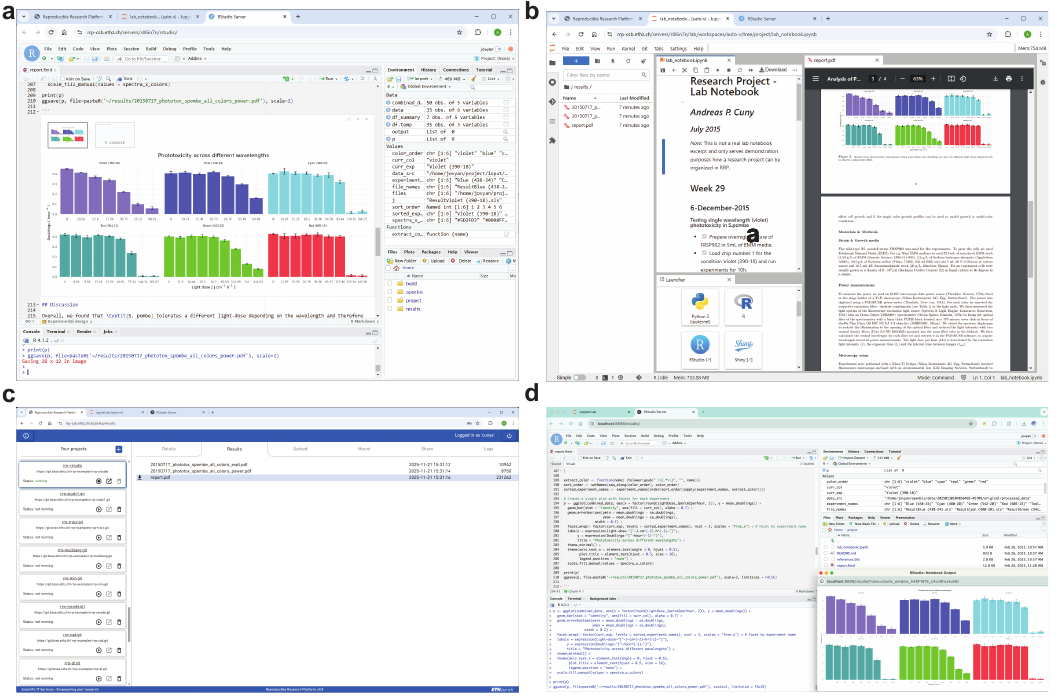}
	\caption{Example of a lab notebook using the R programming language and R-Studio for analysis and report generation. \textbf{a}, R-Studio interface launched from RRP allows users to keep their habits and work with R Markdown to generate analysis and reports. \textbf{b}, The JupyterLab interface of the RRP project with an R notebook and view of the report PDF generated with R-Studio. \textbf{c}, Results generated in the RRP projects are accessible from the UI directly for download or registration in the LIMS (Upload tab). \textbf{d}, Sharing of the R-project with colleagues allows reproducing the results on a local machine running a different operating system (macOS) after downloading the shared "player bundle".}
	\label{fig.rrp_si_r_usecase}
\end{extendeddatafigure}

\begin{extendeddatafigure}[ht]
	\centering
      \includegraphics[width=16cm ]{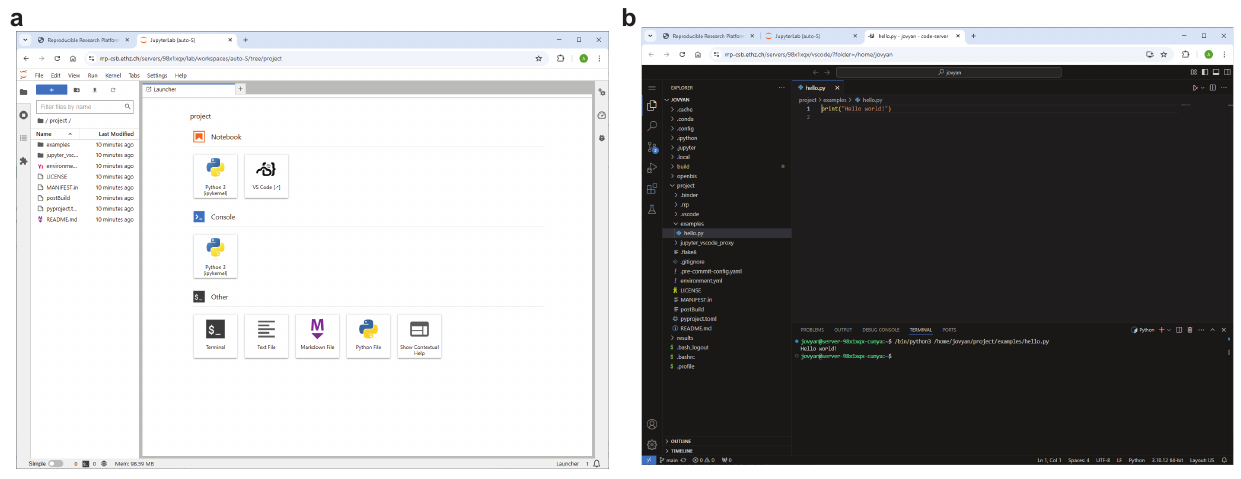}
	\caption{RRP allows using integrated development environments (IDE). \textbf{a}, RRP project with Visual Studio Code (VS Code) IDE installed. \textbf{b}, Simple Python script "hello\_world.py" run from within VS Code IDE on RRP.}
	\label{fig.rrp_si_vscode}
\end{extendeddatafigure}

\begin{extendeddatafigure}[ht]
	\centering
      \includegraphics[width=16cm ]{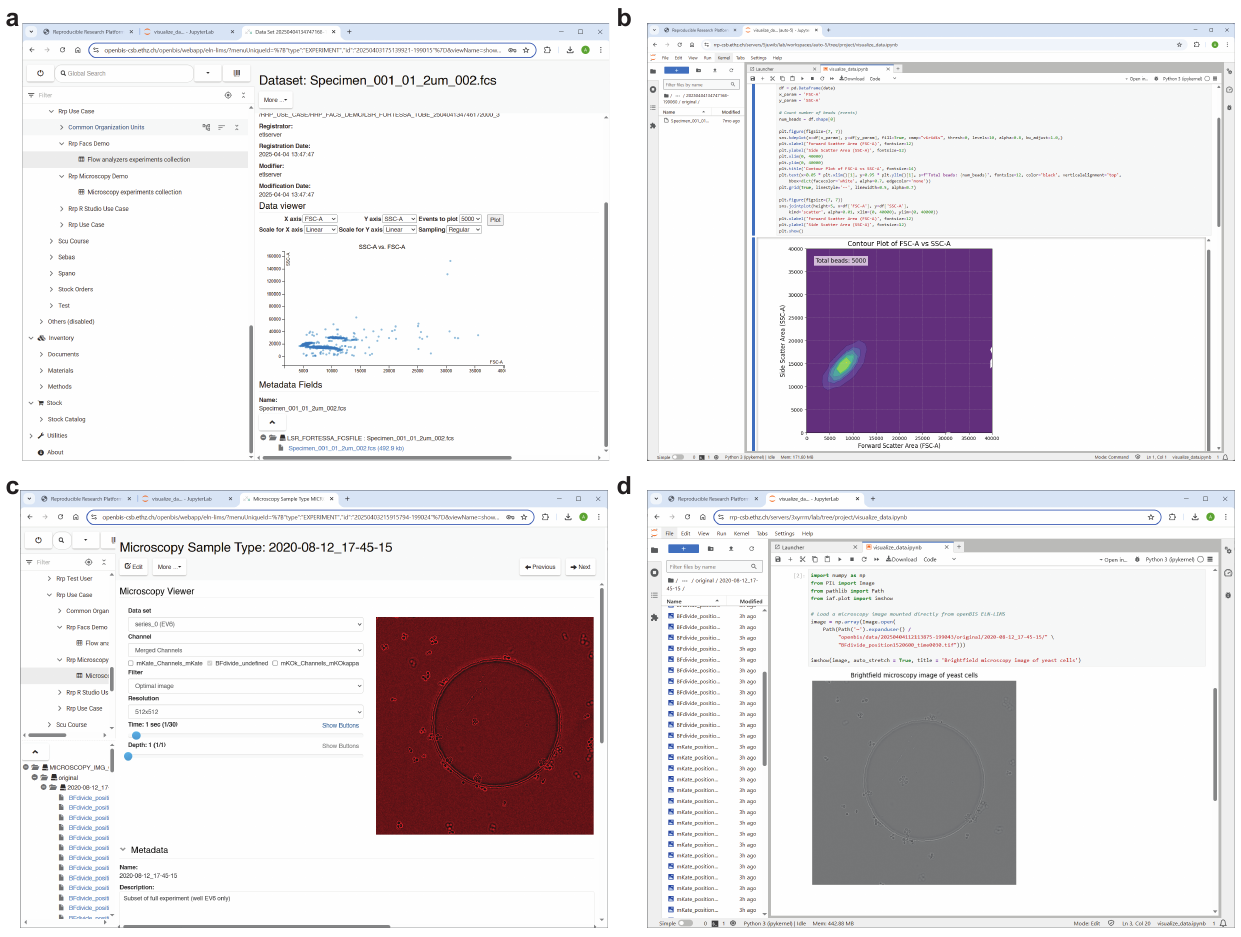}
	\caption{Examples for the interaction with microscopy and Fluorescence Activated Cell Sorting (FACS) data in the RDMS openBIS and when mounted in RRP. \textbf{a}, openBIS ELN-LIMS view of FACS data allows for direct inspection and management of metadata. \textbf{b}, The same FACS data set as in (\textbf{a}) mounted in RRP allows for direct reproducible processing such as gating and analysis with user-defined software tools. \textbf{c}, OpenBIS ELN-LIMS view of microscopy data gives a quick overview of e.g., a time-lapse experiment along the experiment's metadata. \textbf{d}, When mounted in RRP, big data is directly available for further processing and analysis, overcoming the need to move large data sets around (e.g., download). The raw data is read-only to prevent data loss.}
	\label{fig.rrp_si_mic_facs}
\end{extendeddatafigure}

\end{extendeddatafiguresenv}

\newpage\clearpage
\section{Supplementary Information}
\beginsupplement

\begin{table}[ht]
\centering
\small
\begin{tabular}{lp{6.3cm}l}
\hline
\textbf{Title}          & \textbf{Link} & \textbf{DOI}                                                             \\ \hline
RRP platform source code            & \href{https://sissource.ethz.ch/sispub/reproducible-research-platform.git}{sispub/reproducible-research-platform.git}                 &  n/a    \\ \hline
RRP platform demo              & \href{https://sissource.ethz.ch/sispub/reproducible-research-platform-demo.git}{sispub/reproducible-research-platform-demo.git}            &  n/a        \\ \hline
RRP project demo             & \href{https://sissource.ethz.ch/sispub/reproducible-research-platform-demo-project.git}{sispub/reproducible-research-platform-demo-project.git}            & \href{https://doi.org/10.5281/zenodo.17421726}{10.5281/zenodo.17421726}         \\ \hline
RRP minimal template     & \href{https://git.bsse.ethz.ch/rrp-examples/rrp-min.git}{rrp-examples/rrp-min.git}               &  \href{https://doi.org/10.5281/zenodo.17376398}{10.5281/zenodo.17376398} \\ \hline
\\ \hline
\end{tabular}
\caption{Table of RRP source code, demo code and related RRP demo git project and a RRP minimal template as a cookie-cutter.}
\label{table.rrp_source_table}
\end{table}

\begin{table}[ht]
\centering
\small
\begin{tabular}{lll}
\hline
\textbf{Title}          & \textbf{Link} & \textbf{DOI}                                                             \\ \hline
RRP Use-case            & \href{https://git.bsse.ethz.ch/rrp-examples/rrp-use-case}{rrp-examples/rrp-use-case.git}             &  \href{https://doi.org/10.5281/zenodo.17356172}{10.5281/zenodo.17356172}        \\ \hline
RRP Study 1            & \href{https://git.bsse.ethz.ch/rrp-examples/rrp-study1}{rrp-examples/rrp-study1.git} & \href{https://doi.org/10.5281/zenodo.17396968}{10.5281/zenodo.17396968} \\ \hline
RRP Study 2            & \href{https://git.bsse.ethz.ch/rrp-examples/rrp-study2}{rrp-examples/rrp-study2.git} & \href{https://doi.org/10.5281/zenodo.17432259}{10.5281/zenodo.17432259} \\ \hline
RRP VS Code              & \href{https://git.bsse.ethz.ch/rrp-examples/rrp-vscode.git}{rrp-examples/rrp-vscode.git}                     & \href{https://doi.org/10.5281/zenodo.17404794}{10.5281/zenodo.17404794} \\ \hline
RRP RStudio             & \href{https://git.bsse.ethz.ch/rrp-examples/rrp-rstudio.git}{rrp-examples/rrp-rstudio.git}                    & \href{https://doi.org/10.5281/zenodo.17395599}{10.5281/zenodo.17395599} \\  \hline
RRP LaTeX               & \href{https://git.bsse.ethz.ch/rrp-examples/rrp-latex.git}{rrp-examples/rrp-latex.git}              & \href{https://doi.org/10.5281/zenodo.17311271}{10.5281/zenodo.17311271} \\  \hline
RRP Desktop             & \href{https://git.bsse.ethz.ch/rrp-examples/rrp-desktop.git}{rrp-examples/rrp-desktop.git} & \href{https://doi.org/10.5281/zenodo.17378403}{10.5281/zenodo.17378403}\\ \hline
RRP Genetic Engineering & \href{https://git.bsse.ethz.ch/rrp-examples/rrp-genetic-engineering.git}{rrp-examples/rrp-genetic-engineering.git} & \href{https://doi.org/10.5281/zenodo.17394998}{10.5281/zenodo.17394998} \\  \hline
RRP PCB                 & \href{https://git.bsse.ethz.ch/rrp-examples/rrp-pcb.git}{rrp-examples/rrp-pcb.git}                & \href{https://doi.org/10.5281/zenodo.17376573}{10.5281/zenodo.17376573} \\  \hline
RRP DL                  & \href{https://git.bsse.ethz.ch/rrp-examples/rrp-dl.git}{rrp-examples/rrp-dl.git }                & \href{https://doi.org/10.5281/zenodo.17407978}{10.5281/zenodo.17407978} \\  \hline
RRP CAD                 & \href{https://git.bsse.ethz.ch/rrp-examples/rrp-cad.git}{rrp-examples/rrp-cad.git} & \href{https://doi.org/10.5281/zenodo.17376745}{10.5281/zenodo.17376745} \\ \hline
RRP Microscopy and FACS & \href{https://git.bsse.ethz.ch/rrp-examples/rrp-mic-facs.git}{rrp-examples/rrp-mic-facs.git} & \href{https://doi.org/10.5281/zenodo.17394691}{10.5281/zenodo.17394691} \\  \hline
RRP Matlab with TracX & \href{https://git.bsse.ethz.ch/rrp-examples/rrp-tracx.git}{rrp-examples/rrp-tracx.git} & \href{https://doi.org/10.5281/zenodo.17432841}{10.5281/zenodo.17432841} \\  \hline
RRP Multi-Language      & \href{https://git.bsse.ethz.ch/rrp-examples/rrp-multilang.git}{rrp-examples/rrp-multilang.git} & \href{https://doi.org/10.5281/zenodo.17413300}{10.5281/zenodo.17413300}\\  \hline
\end{tabular}
\caption{Table of all RRP projects used in this study with DOI of their player bundles.}
\label{table.rrp_projects_table}
\end{table}

\end{document}
\typeout{get arXiv to do 4 passes: Label(s) may have changed. Rerun}